\documentclass[preprint,prb,showpacs]{revtex4}

\usepackage{graphicx}
\usepackage{dcolumn}
\usepackage{bm}
\usepackage{natbib}

\begin{document}

\title{\boldmath Electron-hole symmetry in quasiparticle spectral weight of cuprates observed via infrared and photoemission spectroscopy \unboldmath}

\author{Myounghoon Lee$^1$}\thanks{These authors contributed equally.}
\author{Dongjoon Song$^{2,3}$}\thanks{These authors contributed equally.}
 \author{Yu-Seong Seo$^1$} \author{Seulki Roh$^1$} \author{Seokbae Lee$^1$} \author{Hiroshi Eisaki$^3$} \author{Jungseek Hwang$^{1}$}\email{jungseek@skku.edu}
\affiliation{$^1$Department of Physics, Sungkyunkwan University, Suwon, Gyeonggi-do 16419, Republic of Korea\\ $^2$Department of Physics and Astronomy, Seoul National University, Seoul 08826, Republic of Korea\\ $^3$National Institute of Advanced Industrial Science and Technology, Tsukuba 305-8568, Japan}

\date{\today}

\begin{abstract}
We performed an optical spectroscopy study on single crystals of Pr$_{0.85}$LaCe$_{0.15}$CuO$_{4-\delta}$ (PLCCO) to revisit the electron-hole asymmetry, which has been understood as a fundamental property of cuprates. Four differently annealed samples – as-grown, reduced, optimally oxygenated, and over-oxygenated samples – were prepared, which have superconducting transition temperatures, $T_c$ = 0, 15, 24, and 18 K, respectively. We observed that low-energy quasiparticle spectral weights of all the PLCCO samples are significantly small in comparison with those of other electron-doped cuprate families. Instead, they are rather close to those of hole-doped counterpart La$_{2-x}$Sr$_x$CuO$_4$ (LSCO). Accordingly, estimated effective carrier numbers per Cu atom ($N_{\mathrm{eff}}$/Cu) of superconducting samples are also considerably small, despite their relatively high critical temperatures. Complementary photoemission study reveals that the low-energy quasiparticle spectral weight of PLCCO is much smaller than that of Nd$_{1.85}$Ce$_{0.15}$CuO$_{4-\delta}$ (NCCO), consistent with the optical results. Our observations demonstrate that PLCCO provides the electron-hole symmetry in quasiparticle spectral weight, and highlight the importance of Cu3$d$–O2$p$ hybridization to understand the low-energy spectral weight transfer in doped cuprates.

\end{abstract}
\pacs{74.25.Gz,74.20.Mn,74.25.-q}

\maketitle

\section{Introduction}
Since the discovery of high-temperature superconducting copper oxides (or cuprates) in 1986 \cite{bednorz:1986}, extensive experimental efforts have established asymmetric phase diagrams between electron- and hole-doped cuprate systems by comparison of rare-earth cuprates, $RE_2$CuO$_4$ with $RE$ = Nd with the T’ structure and La with the T structure for electron- and hole-doped sides, respectively \cite{tagagi:1989,cooper:1990,arima:1993,fujita:2003,onose:2004,wang:2006,homes:2006a,armitage:2010,ohnishi:2018,fournier:2015}.  Prototypically, the electron-doped side exhibits a wider antiferromagnetic (AF) phase and narrower and smaller superconducting (SC) dome than the hole-doped side. However, considerable studies on the phase diagrams of the electron-doped cuprates showed that the phase diagrams are significantly dependent upon rare-earth elements. For example, as the $RE$ element varies from La, Pr, Nd, Sm, Eu, and Gd, superconductivity becomes weaker and eventually completely suppressed \cite{armitage:2010,krockenberger:2008}. Because $RE$ element-dependent studies have not been firmly established, the electron-hole asymmetry as a universal feature of cuprates is still under debate.

In recent years, the long-standing conjecture of the electron-hole asymmetry has been challenged as Nd for the $RE$ site of electron-doped cuprates is replaced by La and Pr. Indeed, electron-doped La$_{2-x}$Ce$_x$CuO$_4$ and Pr$_{1-x}$LaCe$_x$CuO$_4$ (PLCCO) exhibited a similar phase diagram with the archetypal hole-doped cuprate La$_{2-x}$Sr$_x$CuO$_4$ (LSCO) \cite{krockenberger:2008,song:2017}. Moreover, phase diagrams with narrower AF phase and wider superconducting dome compared with LSCO have been reported in both bulk and thin-film electron-doped cuprate systems \cite{krockenberger:2013,horio:2016}, which contradicts the existing notion. Therefore, revisiting the electron-hole asymmetry by substantiating other properties of cuprates is highly encouraging to provide fundamental insights into the origins of high-temperature superconductivity and other novelties in cuprates.

From a perspective of electrodynamics, one characteristic distinguishing electron doping from hole one is the asymmetry in the optical quasiparticle spectral weight or effective carrier number per Cu atom ($N_{\mathrm{eff}}$/Cu) \cite{millis:2004,comanac:2008,ohnishi:2018}. Previous optical spectroscopic studies showed that in comparison to the hole-doped LSCO, $N_{\mathrm{eff}}$/Cu increases more rapidly with doping in the electron-doped Nd$_{2-x}$Ce$_x$CuO$_{4-\delta}$ (NCCO), indicating easy metallization of cuprates with electron-doping \cite{comanac:2008,ohnishi:2018,weber:2010}. On the other hand, twice more $N_{\mathrm{eff}}$/Cu is required to induce superconductivity in NCCO than in LSCO, which strengthens the prevailing prejudice that electron-doped cuprates are less efficient superconductors \cite{comanac:2008,ohnishi:2018,weber:2010}. Therefore, the possible electron-hole symmetry between the phase diagrams of PLCCO and LSCO calls for further investigations on the quasiparticle spectral weight by using spectroscopic experimental techniques such as optical and photoemission techniques.

In this paper, we performed the optical spectroscopy study on the electron-doped cuprate PLCCO (with $x$ =0.15) to revisit the electron-hole asymmetry in $N_{\mathrm{eff}}$/Cu. We measured the reflectance spectra of four differently annealed PLCCO at various temperatures and obtained the optical conductivity from the measured reflectance spectra. Integrating the optical conductivity up to cutoff frequency ($\sim$ 1.12 eV), we found that PLCCO shows significantly lower $N_{\mathrm{eff}}$/Cu than Pr$_{2-x}$Ce$_x$CuO$_{4-\delta}$ (PCCO) and NCCO with a similar $T_c$. Instead, $N_{\mathrm{eff}}$/Cu of our electron-doped PLCCO samples seem to be closer to those of hole-doped LSCO, which might be related to the particle-hole symmetric phase diagram of PLCCO and LSCO. A complementary angle-resolved photoemission spectroscopy (ARPES) study yields consistent results that the quasiparticle spectral weight of PLCCO is smaller than that of NCCO. Based on a recent theoretical result \cite{jang:2016}, we can elucidate that the longest Cu-O bonding length of PLCCO among the bulk electron-doped families yields the largest effective charge-transfer energy, possibly resulting in small spectral weight transfer to the quasiparticle state.

\section{Experiments}

The high-quality PLCCO and NCCO single-crystals were prepared by a traveling solvent floating zone method followed by reduction and oxygenation annealing processes. The nominal Ce concentration for both samples is 15\% ($x$ = 0.15). More detailed annealing procedures and characteristic properties of the samples have been introduced in Ref. [\cite{song:2017}]. The electron concentrations ($n$) of reduced, optimally oxygenated, and over-oxygenated samples of PLCCO determined by ARPES are around 0.11, 0.14, and 0.17, respectively \cite{song:2017}. The superconducting transition temperatures ($T_c$) of the annealed samples are 15 K ($n$ = 0.11), 24 K ($n$ = 0.14), and 18 K ($n$ = 0.17); the sample with $n$ = 0.14 is nearly optimally doped (see Supplementary Information). The as-grown sample does not exhibit superconductivity. All the samples were cut and polished along the $ab$ plane for the optical study. The areas of the samples were roughly 3$\times$3 mm$^2$. The Laue patterns of the cut and polished surfaces of the samples are measured and the miscut angles are obtained from the Laue measurements (see Supplementary Information). They are roughly 4$^{\circ}$, 2$^{\circ}$, 5$^{\circ}$, and 2$^{\circ}$ for the as-grown, $n$ = 0.11, 0.14, and 0.17 samples, respectively.

To investigate temperature-dependent optical properties of PLCCO, we measured the $ab$ plane reflectance of each sample across a wide spectral range from 75 to 35000 cm$^{-1}$ at various selected temperatures ranging from 8 to 300 K (see Supplementary Information). For the reflectance measurements, the incident angle was near normal ($\sim$11$^{\circ}$). To cover the wide spectral range, a Fourier-transform infrared (FTIR)-type spectrometer (Bruker Vertex 80v: 75 – 8000 cm$^{-1}$) and a monochromatic spectrometer (Perkin-Elmer Lambda 950: 5000 – 35000 cm$^{-1}$) were used. For the temperature-dependent study, a continuous flow liquid He cryostat was used. An {\it in-situ} metallization method \cite{homes:1993} was employed to obtain accurate reflectance spectra. We checked how accurately the samples were cut and polished along the $ab$ plane with a simple experiment using a polarizer and confirmed that miscut effects in the measured reflectance spectra were negligibly small (see Supplementary Information). The optical conductivity was obtained from the measured reflectance spectrum by using the Kramers–Kronig (KK) relation, Fresnel formula, and relations between optical constants \cite{wooten,tanner:2019}. To perform the KK analysis, one has to extrapolate the measured reflectance spectrum in a finite spectral range to both zero and infinity frequencies. For the extrapolation from the lowest data point to zero frequency, the Hagen–Rubens relation (i.e., $1 - R(\omega) \propto \sqrt{\omega}$) and $1 - R(\omega) \propto \omega^4$ were used for the normal and superconducting states, respectively. For the extrapolation from the highest data point (35,000 cm$^{-1}$) to the infinity, it was assumed that $R(\omega) \propto 1/\omega$ up to 1.0$\times$10$^6$ cm$^{-1}$ and the reflectance follows the free electron behavior\cite{wooten,tanner:2019} (i.e., $R(\omega) \propto 1/\omega^4$) above 1.0$\times$10$^6$ cm$^{-1}$.

ARPES experiments on PLCCO and NCCO were carried out at beamline 5-2 in Stanford Synchrotron Radiation Lightsource (SSRL). The samples were cleaved in-situ in a ultra-high vacuum lower than 4$\times$10$^{-11}$ Torr. The measurement temperature was around 10 K. Linearly polarized photons with various energies were used to quantitatively make sure that the spectral weight of quasiparticle band was relative to that of high binding energy bands.

\section{Results and discussion}

\begin{figure}[!htbp]
  \vspace*{-0.3 cm}%
  \centerline{\includegraphics[width=5.0 in]{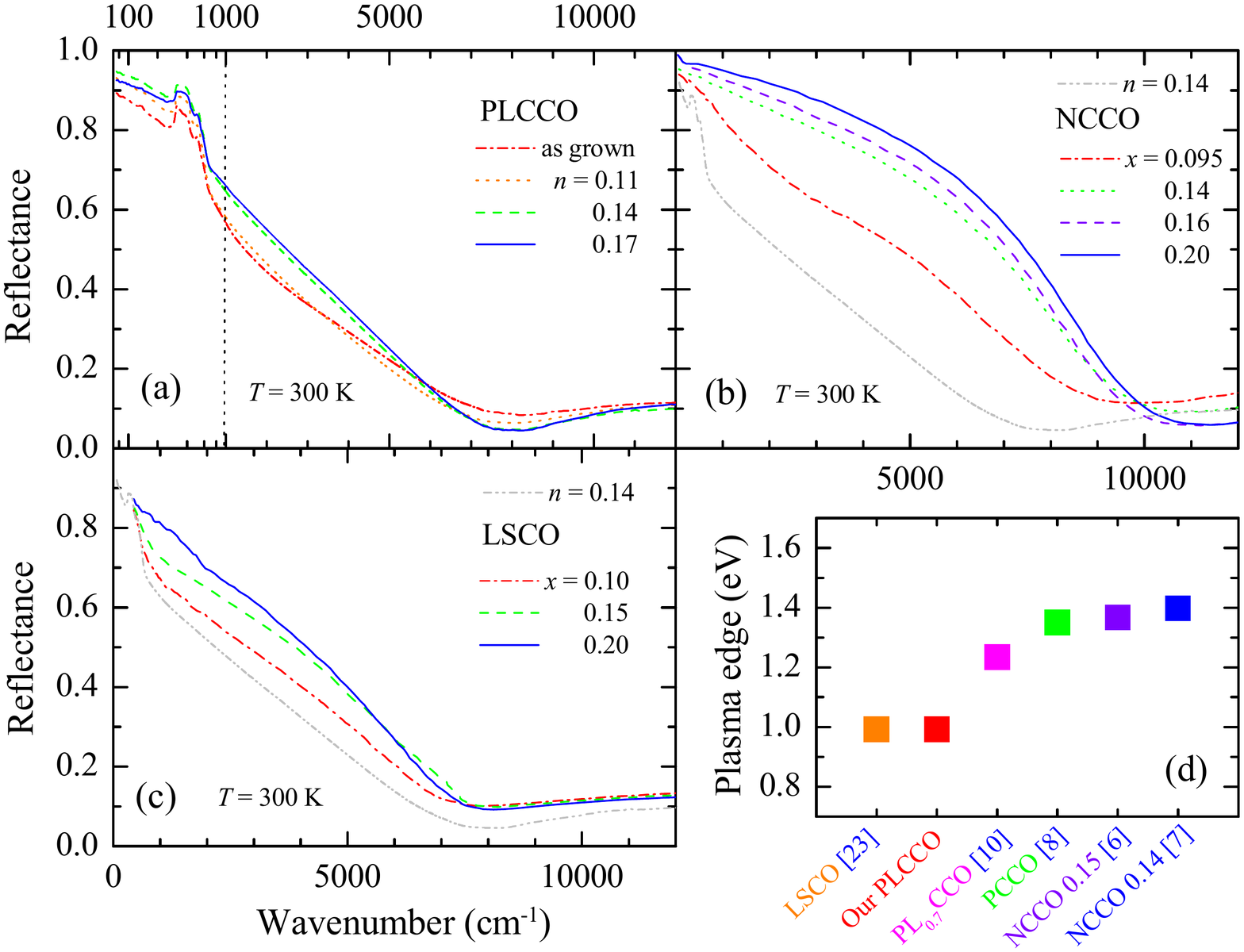}}%
  \vspace*{-0.3 cm}%
\caption{(Color online) Comparison of reflectance spectra of PLCCO and those of NCCO and LSCO at 300 K. (a) Electron-concentration (or doping) dependent reflectance spectra of the four PLCCO samples at $T$ = 300 K. We note that, on the horizontal axis, we plot the spectra up to 1000 cm$^{-1}$ on a long scale and above it on a linear scale. (b) Reproduced reflectance spectra of electron-doped NCCO \cite{wang:2006} at 300 K along with the reflectance spectrum of our PLCCO sample with $n$ =0.14 at the same temperature. (c) Reproduced reflectance spectra of hole-doped LSCO\cite{uchida:1991} at 300 K along with the reflectance spectrum of our PLCCO sample with $n$ = 0.14 at the same temperature. (d) Plasma edges of several cuprate systems at near optimally doping level.}
 \label{fig1}
\end{figure}

Measured reflectance spectra of the four PLCCO samples at various temperatures can be found in Supplementary Information. We compared the measured reflectance spectra of PLCCO with those of electron-doped NCCO \cite{onose:2004,wang:2006} and PCCO \cite{arima:1993,homes:2006a}, and hole-doped LSCO \cite{uchida:1991} previously reported in literature. We display the measured reflectance spectra of the four PLCCO samples at 300 K in Figure 1(a). We observe a linear frequency dependence of reflectance spectra (also see Fig. 2(c) and (d)) in a wide spectral range for the optimally doped and overdoped samples, which is suggestive of a marginal Fermi liquid response \cite{varma:1989,hwang:2004a}. The plasma edge appears as a dip near 8000 cm$^{-1}$. We note that the plasma edge of the as-grown sample is less well-defined, which may be related to the ARPES observation that the Fermi surface of this sample is not well-defined \cite{song:2017}. The concentration-dependent (or heat-treatment-dependent) reflectance change can be clearly seen in Fig. 1(a). The reflectance in the mid-infrared (MIR) region below 6000 cm$^{-1}$ increases with the concentration, which is a common doping-dependent behavior in cuprate systems. However, the magnitude of the doping-dependent reflectance change is much smaller than those of other electron-doped cuprates as observed in Fig. 1(b). This can be associated with the widest superconducting dome of PLCCO among the electron-dope families \cite{lin:2021}. To further study the different doping-dependent reflectance changes, we display the reflectance spectra of electron-doped NCCO \cite{wang:2006} and hole-doped LSCO \cite{uchida:1991} at various doping levels at $T$ = 300 K in Fig. 1(b) and 1(c), respectively, along with the reflectance of our PLCCO sample with $n$ = 0.14 at the same temperature. In Fig. 1(b), the reflectance of PLCCO at $n$ = 0.14 is lower than that of NCCO at $x$ = 0.095. We can see also that the doping-dependent reflectance change in NCCO is much larger than that in PLCCO. The plasma edge of PLCCO ($\sim$8000 cm$^{-1}$) is lower than those of NCCO (9000 - 11,000 cm$^{-1}$).

On the other hand, in Fig. 1(c), we observe that the reflectance of PLCCO at $n$ = 0.14 is rather comparable with that of LSCO at $x$ = 0.10. Moreover, the magnitude of the doping-dependent change of LSCO is similar to that of PLCCO as can be seen in Fig 1(a). The plasma edges of PLCCO and LSCO are similar as well. In Fig. 1(d), we show the plasma edges of several electron-doped cuprate systems \cite{arima:1993,onose:2004,wang:2006,homes:2006a,ohnishi:2018,uchida:1991}, including a hole-doped LSCO system. Because the plasma edge is intimately associated with the charge carrier density we expect that our PLCCO samples may have a similar charge carrier density to LSCO that is lower than the charge densities of other electron-doped cuprates.

\begin{figure}[!htbp]
  \vspace*{-0.3 cm}%
  \centerline{\includegraphics[width=4.7 in]{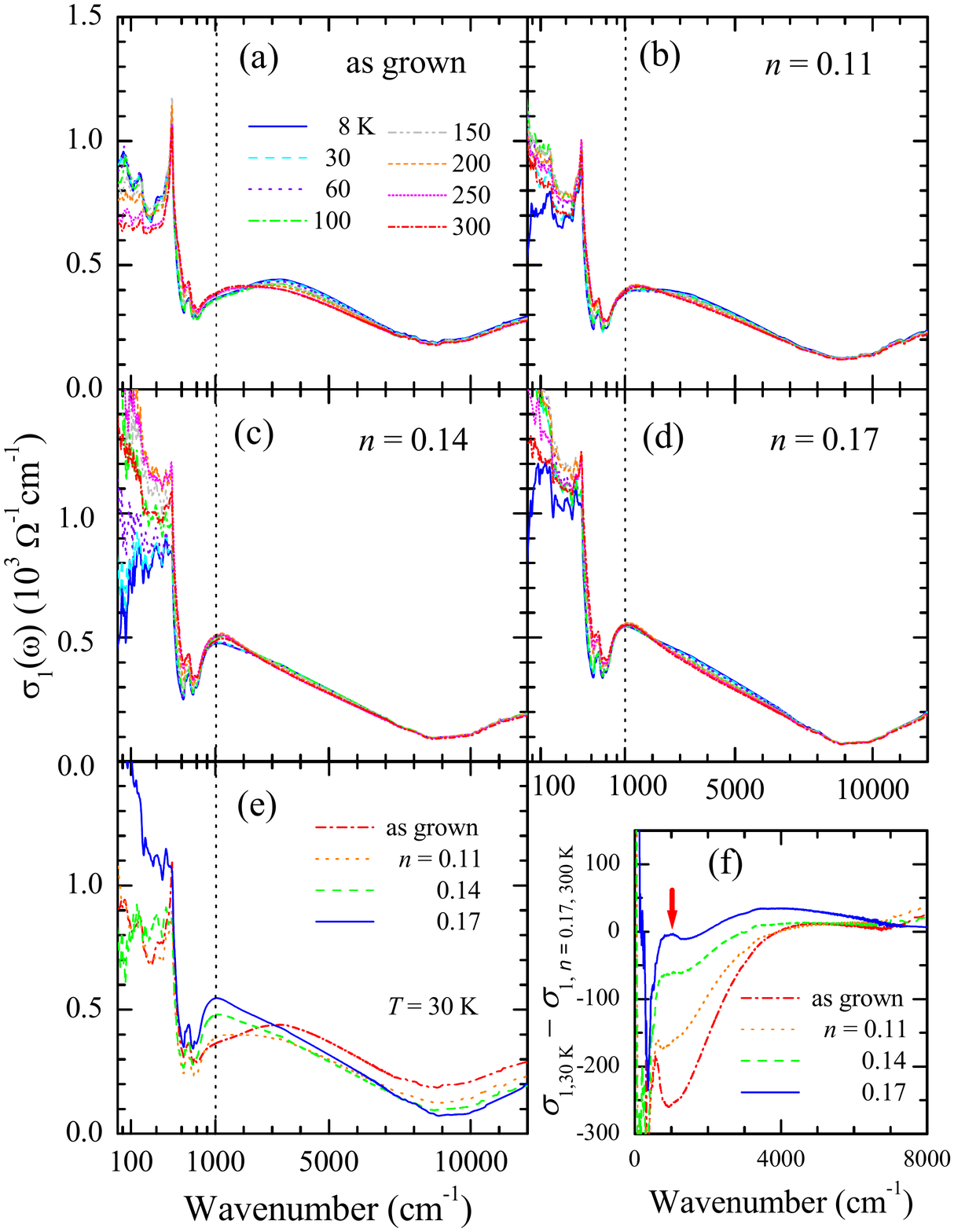}}%
  \vspace*{-1.3 cm}%
\caption{(Color online) Optical conductivity spectra of four PLCCO samples at various selected temperatures ranging from 8 to 300 K: (a) as grown, (b) $n$ = 0.10, (c) $n$ = 0.14, and (d) $n$ = 0.17. (e) Optical conductivity spectra of the four PLCCO samples at $T$ = 30 K. It is worth noting that, on the horizontal axes of (a-e), we plot the spectra up to 1000 cm$^{-1}$ on a long scale and above it on a linear scale. (f) Difference between the optical conductivity at $T$ = 30 K and that of the highest doped sample with $n$ =0.17 at 300 K. The vertical arrow indicates the pseudogap feature, which is closely associated with the pseudogap at the hot spot in APRES spectrum \cite{park:2007}.}
 \label{fig2}
\end{figure}

Fig. 2(a) - (d) show the optical conductivities of the four PLCCO samples obtained from the measured reflectance spectra using the KK analysis. Owing to the low reflectance of the four PLCCO samples, the overall levels of the optical conductivity spectra are significantly lower than those of other electron-doped cuprates \cite{arima:1993,onose:2004}; the average optical conductivity is one-third that of similarly doped electron-doped cuprates. Interestingly, the overall levels of conductivity are comparable with those of a hole-doped cuprate, underdoped LSCO system\cite{startseva:1999}. We plot the optical conductivity spectra in the low frequency region in log (70-1000 cm$^{-1}$) and linear (1000-12000 cm$^{-1}$) scales in the horizontal axes of Fig. 2(a)-(e) to better illustrate them. All four samples show a prominent sharp peak near 300 cm$^{-1}$, which is the same phonon previously observed in the as-grown NCCO system. As the doping increases, the spectral weight in the low frequency region increases, and the phonon is screened by the enhanced spectral weight. The peak is followed by a dip on the higher frequency side, which looks like an antiresonance\cite{singley:2001}. The dip persists in the overdoped sample ($n =$ 0.17). However, for the NCCO system, the dip is almost completely suppressed when the sample becomes a superconductor by deoxygenation\cite{singley:2001}. We speculate that the dip feature might be associated with the additional La atom in our samples because the dip feature was observed in the optical conductivity of reduced PL$_{0.7}$CCO systems\cite{ohnishi:2018}. We note that because we do not have spectra below 75 cm$^{-1}$, which is larger than the superconducting gap ($2\Delta_s \simeq$ 32 cm$^{-1}$)\cite{armitage:2001}, we decided not to discuss superconducting properties of PLCCO such as the missing spectral weight and the superfluid density.

All samples exhibit spectral weight redistributions in the MIR region as the temperature changes. As temperature drops, the spectral weight transfers from low to high frequency, resulting in resulting in a partial suppression (or gap) in the spectrum by shifting a broad peak near 1000 cm$^{-1}$ to near 3500 cm$^{-1}$. Therefore, a spectral weight loss at low frequency is recovered at high frequency. It is worth noting that as doping increases, the spectral weight redistribution becomes weaker, the broad peak near 1000 cm$^{-1}$ at 300 K shifts to a low frequency, and the broad peak near 3500 cm$^{-1}$ at 30 K becomes less well-defined. Interestingly, the same kind of spectral weight transfer occurs as decreasing the doping, as shown in Fig. 2(e), where we display the optical conductivity spectra of the four samples at 30 K. The optical conductivity of the as-grown sample shows a broad peak near 3500 cm$^{-1}$. As the doping increases, the broad peak becomes suppressed and the spectral weight is released to the low frequency region resulting in appearance of a peak near 1000 cm$^{-1}$. This feature appears with consuming the spectral weight in low frequency region. This doping-dependent pseudogap behavior is very similar to the temperature-dependent one (compare Fig. 2(a)). These temperature- and doping-dependent spectral weight redistributions are closely related to the antiferromagnetic pseudogap features observed in underdoped NCCO system\cite{onose:2004}. To show the doping-dependent pseudogap feature more clearly, in Fig. 2(f), we show the difference spectra between the optical conductivity at $T$ = 30 K and that of the highest doped sample with $n$ =0.17 at 300 K, which shows the weakest pseudogap effect because the pseudogap may close with an increase in either the doping or the temperature. The red vertical arrow indicates a suppression in the optical conductivity, which can be explained as a pseudogap feature that appears near the hot spot in the ARPES spectrum \cite{park:2007}.

\begin{figure}[!htbp]
  \vspace*{-0.3 cm}%
  \centerline{\includegraphics[width=5.5 in]{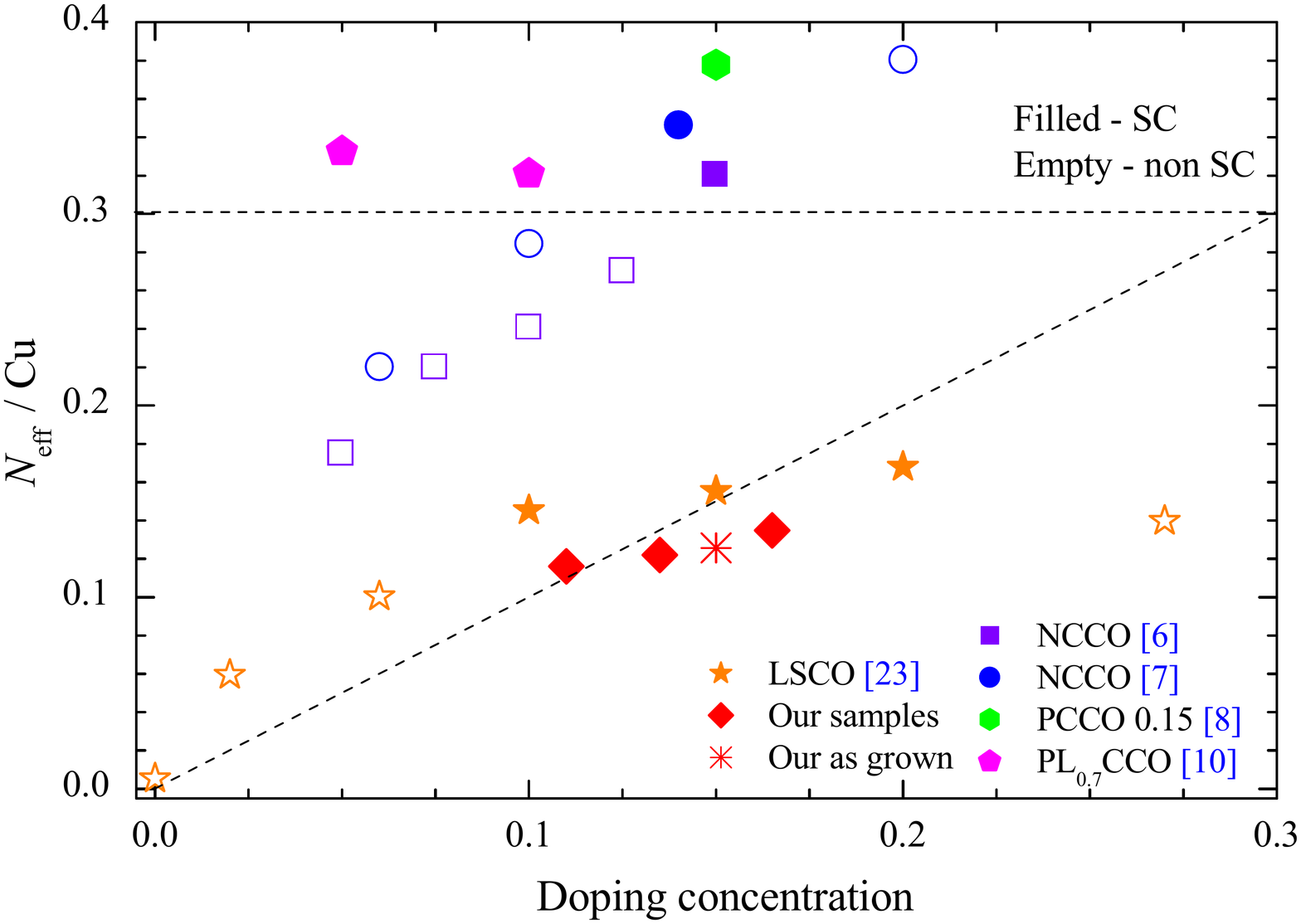}}%
  \vspace*{-1.0 cm}%
\caption{(Color online) Calculated effective electron number per Cu atom ($N_{\mathrm{eff}}$/Cu) of several cuprate systems. The superconducting and normal states are presented by filled and empty symbols, respectively. The thin dashed diagonal line shows the linear relation between $N_{\mathrm{eff}}$/Cu and doping concentration. We note that all $N_{\mathrm{eff}}$/Cu were estimated from spectra at a normal state except for one sample of NCCO \cite{onose:2004} with a Ce concentration of 0.15.}
 \label{fig3}
\end{figure}

For quantitative evaluation of the number of charge carriers, we evaluated the effective electron number per Cu atom ($N_{\mathrm{eff}}$/Cu) by using the following formula \cite{ohnishi:2018}:
\begin{equation}\label{}
N_{\mathrm{eff}}\mbox{/Cu}=\frac{120 m_e c^2V_{Cu}}{e^2} \int_0^{\omega_c}\sigma_1(\omega)d\omega,
\end{equation}
where $m_e$ is the bare electron mass, $c$ is the speed of light in vacuum, $V_{Cu}$ is the volume per Cu atom, $e$ is the elementary charge, and $\omega_c$ is a cutoff frequency. Here, all the frequencies are in cm$^{-1}$ units. The cutoff frequency can be determined for including all the charge carriers. A suggestive cutoff frequency is a frequency, where a broad dip appears in the optical conductivity. Below 75 cm$^{-1}$, which is the lowest data point, we used the Drude model (or the Hagen-Rubens relation) for the extrapolation to zero frequency. For the estimation of the $N_{\mathrm{eff}}$/Cu and better comparison with reported values, we used the bare electron mass instead of the effective mass because accurate estimation of the effective mass from measured optical spectra can be another issue, and other reported values in Ref. \cite{uchida:1991,ohnishi:2018} were estimated by using the bare electron mass instead of the effective mass. We note that other features, such as pseudogap, may redistribute the spectral weight below the interband transition ($\sim$1.12 eV), but they have no effect on the total spectral weight, and the phonon near 300 cm$^{-1}$ has a negligibly small ($<$1\%) contribution to $N_{\mathrm{eff}}$/Cu. Figure 3 shows $N_{\mathrm{eff}}$/Cu of various cuprate systems \cite{cooper:1990,arima:1993,onose:2004,wang:2006,homes:2006a,uchida:1991}, including our PLCCO samples. We note that, for consistent comparisons, we obtain $N_{\mathrm{eff}}$/Cu of LSCO with a cutoff frequency of $\sim$1.12 eV from a previous optical study of LSCO \cite{uchida:1991} and display them in the figure because we use the dip energy ($\sim$1.12 eV) in the optical conductivity as the cutoff frequency. The $N_{\mathrm{eff}}$/Cu of our PLCCO samples are much lower than those of other electron-doped cuprates but they are almost the same as the doping-concentrations estimated by the ARPES study and are roughly proportional to the doping-concentration. We also show $N_{\mathrm{eff}}$/Cu of the as-grown sample (cross symbol), which is similar to those of our treated samples. $N_{\mathrm{eff}}$/Cu of as-grown samples depends on the Ce concentration as reported in a previous optical study \cite{ohnishi:2018}. Interestingly, although the $N_{\mathrm{eff}}$/Cu of our PLCCO samples are well below 0.3 the samples still exhibit superconductivity. This result is a contrast to that of a previous study, which showed that the superconductivity was observed only when the $N_{\mathrm{eff}}$/Cu was near or above 0.3 in Pr$_{1.3-x}$La$_{0.7}$Ce$_x$CuO$_{4+\delta}$ ($x$ = 0.05 and 0.10) samples \cite{ohnishi:2018}. One conclusion, which we can thus draw, is that our PLCCO is a very efficient system for superconductivity in terms of the amount of the electron-concentration; the charge carrier density is much lower compared with other superconducting electron-doped cuprates. It is worth noting that although all four PLCCO samples show small and slightly different spectral weights, the superconducting transition temperatures are clearly tuned by the doping from 0 to 24 K (see Fig. S1). On the other hand, the $N_{\mathrm{eff}}$/Cu of our PLCCO are quite similar to those of hole-doped LSCO \cite{uchida:1991}. This remarkable similarity between PLCCO and LSCO is a reminiscent of their resembled superconducting (or $T_c$) domes against the carrier concentration \cite{song:2017}, suggesting that the main determinants of spectral weight transfer and $T_c$ are correlated with each other. However, it is not clearly understood based on our optical result alone why these 214 families of cuprates with either T' and T structures exhibit the electron-hole symmetric phase diagram.

\begin{figure}[!htbp]
  \vspace*{-0.3 cm}%
  \centerline{\includegraphics[width=4.5 in]{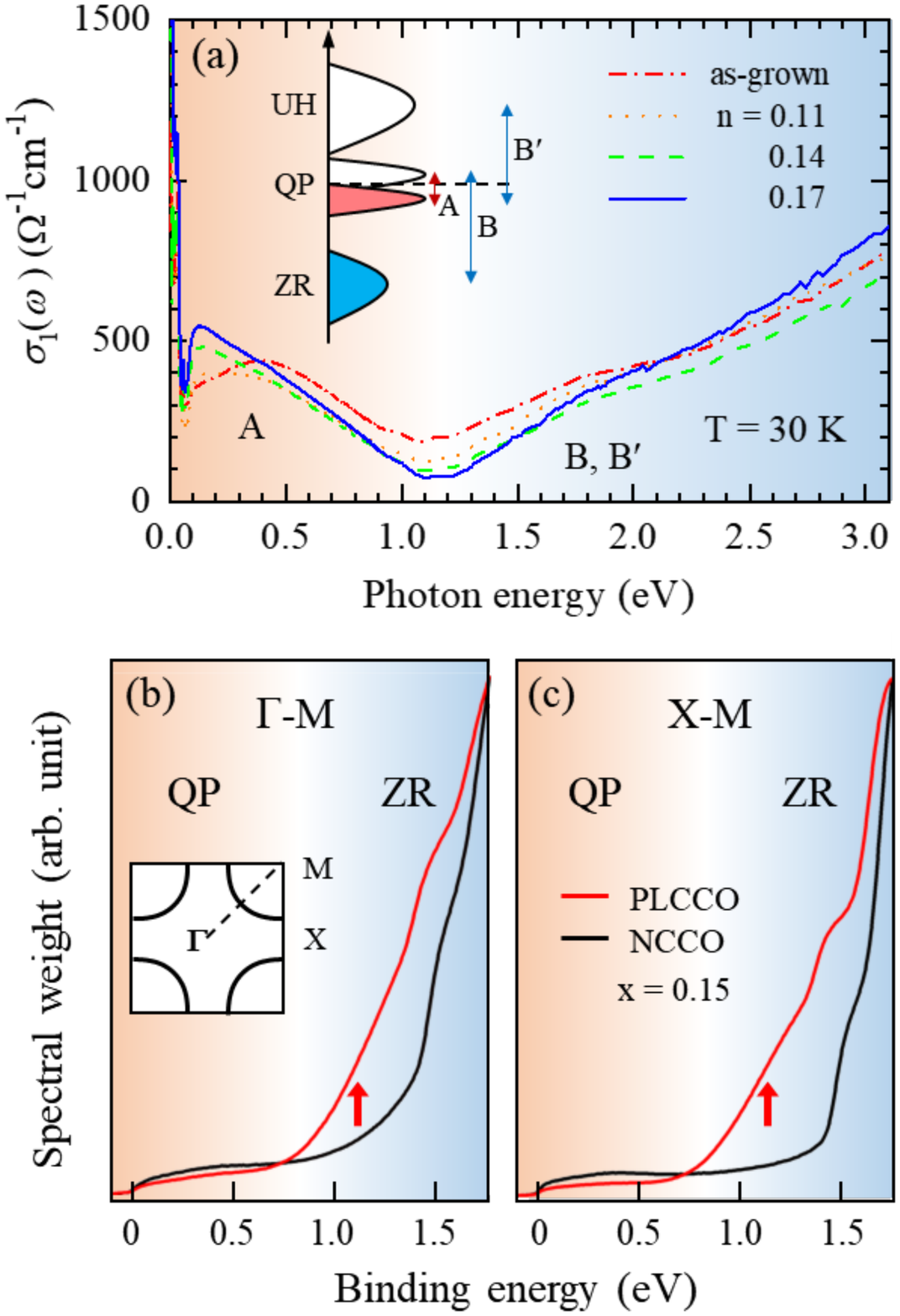}}%
  \vspace*{-0.3 cm}%
\caption{(Color online) Optical conductivity (30 K) and photoemission (10 K) spectra of PLCCO. (a) Optical conductivity with possible interband transitions. UH, QP, and ZR indicate the upper Hubbard band, the quasiparticle band, and the Zhang-Rice band, respectively. (b), (c) Accumulated energy distribution curves of PLCCO and NCCO obtained by integrating the energy dispersions curves of ARPES spectra along $\mathrm{\Gamma}$-M and X-M. Red arrows highlight the high energy band spectra.}
 \label{fig4}
\end{figure}

Figure 4 (a) shows the optical conductivity of PLCCO samples at 30 K with possible inter-band optical transitions (see A, B, and B’) based on the framework of the charge-transfer insulator model with an antiferromagnetically split quasiparticle band \cite{onose:2004,armitage:2010,weber:2010,damascelli:2003,basov:2011a}. The Drude-type and A transitions below ~1.12 eV in the optical conductivity correspond to, respectively, the intra-band transition located at zero energy and the excitations between split quasiparticle bands, while B and B’ above ~1.12 eV are inter-band transitions from Zhang-Rice (ZR) singlet band to quasiparticle (QP) band and from QP to upper-Hubbard (UH) band, respectively. It is worth noting that the two split quasiparticle bands near the Fermi level result from the correlations. Because the $N_{\mathrm{eff}}$/Cu obtained by integration of the optical conductivity below the cutoff energy ($\sim$9,000 cm$^{-1}$ or $\sim$1.12 eV) is a quantity proportional to the quasiparticle spectral weight, the significantly smaller $N_{\mathrm{eff}}$/Cu of PLCCO than NCCO suggests that the spectral weight transfer from those high energy bands to the quasiparticle band in PLCCO is relatively smaller than in NCCO.

To understand the electron-hole symmetry in $N_{\mathrm{eff}}$/Cu between PLCCO and LSCO, we performed further investigations using a complementary experimental technique, ARPES. From comparative photoemission studies between PLCCO and NCCO, we obtained supporting evidence for the smaller spectral weight transfer of PLCCO. Figs. 4(b) and 4(c) show the accumulated energy distribution curves of the PLCCO and NCCO ARPES spectra obtained by integrating the energy dispersion curves along the two high symmetry directions, $\mathrm{\Gamma}$-M (nodal) and X-M (anti-nodal), at 10 K. The nominal Ce content of those two compounds is identical ($x$ = 0.15). They are optimally oxygenated and have similar $T_c$. The ARPES spectra of the QP band for both directions are smaller in PLCCO than in NCCO. On the other hand, the broad spectrum pointed by a red arrow at the energy range of 1.0 – 1.5 eV shows higher intensity in PLCCO, indicating a largely unchanged spectral weight of the ZR band in PLCCO \cite{armitage:2002}. These results are consistent with the small $N_{\mathrm{eff}}$/Cu of PLCCO observed by the optical measurement. It is worth noting that ARPES measures the filled bands while optical spectroscopy measures the transition from filled to empty bands. Therefore, the energy scale of the optical transition is roughly twice that of ARPES. Our combined spectroscopic studies on PLCCO demonstrate that the $N_{\mathrm{eff}}$/Cu of PLCCO is significantly smaller than that of NCCO. Therefore, the $N_{\mathrm{eff}}$/Cu is strongly composition-dependent. Because the $N_{\mathrm{eff}}$/Cu of electron-doped PLCCO is comparable to that of hole-doped LSCO, our results can be counter-evidence against the long-standing belief of electron-hole asymmetry in $N_{\mathrm{eff}}$/Cu.

\section{Discussions}

There has been a controversial issue that the reflectance of electron-doped cuprates shows significant annealing dependence \cite{arima:1993}. Therefore, the present optical results obtained from the differently annealed PLCCO samples manifest that the small $N_{\mathrm{eff}}$/Cu of PLCCO is an intrinsic property of PLCCO irrelevant to the annealing conditions. However, the differently annealed PLCCO samples show different $N_{\mathrm{eff}}$/Cu even though the differences are not very dramatic, as shown in Fig. 3. Electron-doped cuprate PLCCO shows a small $N_{\mathrm{eff}}$/Cu compared with other electron-doped cuprates (PCCO and NCCO), and its small $N_{\mathrm{eff}}$/Cu is comparable to that of LSCO, a hold-doped cuprate. Therefore, the small $N_{\mathrm{eff}}$/Cu or quasiparticle spectral weight in PLCCO may imply that the electron-hole asymmetry in $N_{\mathrm{eff}}$/Cu is not an intrinsic feature of cuprates but depends on material details. We speculate that this observed electron-hole symmetry in $N_{\mathrm{eff}}$/Cu may be intimately associated with the recently reported electron-hole symmetry in the temperature-doping phase diagram\cite{song:2017}.

The fact that the smaller quasiparticle spectral weight results from the larger correlation strength $U$ has been revealed by a previous theoretical study \cite{comanac:2008}. In this picture, the smaller effective $U$ in NCCO than in LSCO yields the larger $N_{\mathrm{eff}}$/Cu in NCCO than in LSCO. Similar scenarios have also been suggested by other theoretical works \cite{ohnishi:2018,weber:2010,mizuno:2017}. Electrostatically, it means that the electron-doped system with the T' structure lacks the apical oxygen, which effectively reduces the correlation strength $U/t$, where $t$ is the hopping parameter \cite{mizuno:2017}. From this perspective, because the longer Cu-O bonding length of PLCCO than NCCO can provide a larger effective correlation strength ($U/t$) because of the smaller $t$, the carriers will be more localized, leading to a smaller $N_{\mathrm{eff}}$/Cu in PLCCO than in NCCO.

Another factor to be considered is the charge-transfer energy $E_d$, the energy deviation between Cu 3$d$ (UH) and O2$p$ (or ZR) energy levels. A previous doping-dependent ARPES study on NCCO manifested the spectral weight transfer from the Cu3$d$-O2$p$ hybridized Zhang-Rice band to the quasiparticle state \cite{armitage:2002}. The study points out that the $N_{\mathrm{eff}}$/Cu of cuprates can be mainly determined by either or both $U/t$ and $E_d/t$. Another recent comparative study on the effective correlation strength $U/t$ and the effective charge-transfer energy $E_d/t$ revealed an intriguing qualitative difference between $U/t$ and $E_d/t$ \cite{jang:2016}. In more detail, as the atomic number of rare-earth in the electron-doped cuprates decreases from Sm to Pr, the $U/t$ decreases whereas the $E_d/t$ increases. As a result, the smaller atomic number makes $U/t$ ($E_d/t$) farther from (closer to) that of La$_2$CuO$_4$, the parent compound of LSCO. Considering that PLCCO has the longest Cu-O bonding length with the smallest rare-earth atomic number among the bulk electron-doped families, the small $N_{\mathrm{eff}}$/Cu with the low cutoff energy of PLCCO, which is comparable with that of LSCO rather than NCCO, suggests that $E_d/t$, instead of $U/t$, is likely the main determinant of the spectral weight transfer in doped cuprates.

Lastly, it should be interesting to note that Nd and Pr have magnetic moments while La does not. Because PLCCO has a relatively higher ratio of the non-magnetic La element than NCCO or PCCO, the formation of a 3-dimensional antiferromagnetic phase is more difficult in PLCCO than in NCCO or PCCO. This difference may suppress the magnetic phase and allow the expansion of both the superconducting and pseudogap phases to the lower doping region in the phase diagram of PLCCO, resulting in the electron-hole symmetric phase diagram. This postulated phase diagram may be intimately related to the high superconductivity efficiency of PLCCO.

Another issue of debate is that the $c$-axis component might suppress the reflectance, resulting in the small $N_{\mathrm{eff}}$/Cu. We carried out reflectance measurements with various angles of an input linear polarizer to observe any significant leakage of the out-of-plane component and confirmed that the measured reflectance spectra mostly came from the in-plane component. Laue X-ray measurements also exhibit well-oriented in-plane diffraction patterns (see Supplementary Information).

\section{Conclusions}

We investigated the PLCCO single crystals using optical spectroscopy combined with the ARPES technique. Our optical study revealed that the overall reflectance spectrum of PLCCO was lower than those of other electron-doped cuprates (NCCO and PCCO) with a similar $T_c$. This lower reflectance reduced the optical conductivity, and consequently, the PLCCO showed a significantly smaller $N_{\mathrm{eff}}$/Cu compared with the other electron-doped families. Instead, we found that the $N_{\mathrm{eff}}$/Cu of PLCCO was rather similar to that of hole-doped LSCO. Our complementary ARPES study also demonstrated a smaller spectral weight transfer from high-biding energy bands to the low energy quasiparticle band in PLCCO than in NCCO. Our observations provide counter-evidence for the electron-hole asymmetry, which has been believed to be a fundamental property of high-$T_c$ cuprates.

% Acknowledgments
%
\acknowledgments This paper was supported by the National Research Foundation of Korea (NRFK Grant No. 2020R1A4A4078780 and 2021R1A2C101109811). This work was supported by the Institute for Basic Science in Korea (Grant No. IBS-R009-G2) ARPES experiments were performed at Beamline 5-2, Stanford Synchrotron Radiation Lightsource, SLAC National Accelerator Laboratory operated by the DOE Office of BES. (Proposal No. 5161)

\bibliographystyle{apsrev4-1}
\bibliography{bib}

\end{document}